# IncentMe: Effective Mechanism Design to Stimulate Crowdsensing Participants with Uncertain Mobility

Francesco Restuccia, *Member, IEEE*, Pierluca Ferraro, Simone Silvestri, *Member, IEEE*, Sajal K. Das, *Fellow, IEEE*, and Giuseppe Lo Re, *Senior Member, IEEE*

**Abstract**—Mobile crowdsensing harnesses the sensing power of modern smartphones to collect and analyze data beyond the scale of what was previously possible. In a mobile crowdsensing system, it is paramount to incentivize smartphone users to provide sensing services in a timely and reliable manner. Given sensed information is often valid for a limited period of time, the capability of smartphone users to execute sensing tasks largely depends on their mobility, which is often uncertain. For this reason, in this paper we propose IncentMe, a framework that solves this fundamental problem by leveraging game-theoretical reverse auction mechanism design. After demonstrating that the proposed problem is NP-hard, we derive two mechanisms that are parallelizable and achieve higher approximation ratio than existing work. IncentMe has been extensively evaluated on a road traffic monitoring application implemented using mobility traces of taxi cabs in San Francisco, Rome, and Beijing. Results demonstrate that the mechanisms in IncentMe outperform prior work by improving the efficiency in recruiting participants by 30%.

**Index Terms**—Mobile, Crowdsensing, Smartphone, Auction, Game Theory, Smartphone, Optimization, Experiments.

✦

## 1 INTRODUCTION

The past few years have witnessed unprecedented proliferation of smartphones in people's daily lives; it has been predicted by Cisco that the total number of smartphones will be nearly 50 percent of global devices and connections by 2020 [1]. Furthermore, nowadays smartphones are equipped with a set of cheap but powerful embedded sensors, such as accelerometer, gyroscope, microphone, and camera.

If all the smartphones on Earth were combined into a single network, it would form the largest sensor network ever created. We could leverage millions of personal smartphones and a near-pervasive wireless network infrastructure to sense, collect, and analyze data far beyond the scale of what was possible before, without the need to deploy thousands of static sensors. This new paradigm is commonly referred as mobile crowdsensing (MCS). Real-world applications of MCS that have been already deployed include monitoring of road traffic [2], environment (e.g., air and noise) [3], [4], crime [5], and parking systems [6].

In order to obtain meaningful data, most of mobile crowdsensing applications need people's active participation to the sensing process. On the other hand, smartphone users invest their personal resources while executing sensing tasks. This implies that a user would not be interested in participating to the crowdsensing process unless she receives a satisfying reward. To this end, the problem of incentive mechanism design in mobile crowdsensing has been extensively studied, as surveyed in [7]–[9]. Briefly, the goal of an incentive mechanism is to optimally (i) schedule sensing tasks to participants; (ii) reward the participants for their services so as to keep them interested in participating.

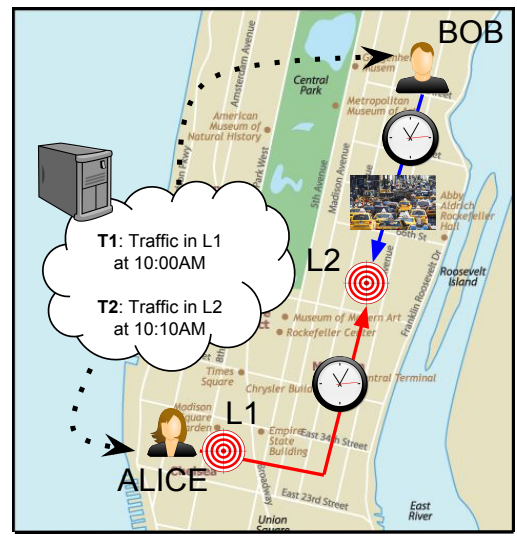

Fig. 1: Example of a crowdsourced navigation app.

**Motivations.** To the best of our knowledge, existing incentive mechanisms implicitly assume that participants will always be able to perform the sensing tasks once assigned to them. On the other hand, participants are most often vehicle drivers, or people commuting to their workplace through public transportation [7]. Thus, participants' mobility largely depends on current traffic conditions and current vehicle speed, implying that they might not always be able to execute sensing tasks despite their willingness to do so.

For this reason, to increase the likelihood of successful execution of the sensing tasks, *the aspect of uncertain mobility should be factored into the incentive mechanism design*.

We now discuss an example to further support this point. Let us consider a (simplified) navigation app with crowd-

- F. Restuccia is with the Department of Electrical and Computer Engineering, Northeastern University, Boston, MA 02215 USA (e-mail: frestuc@ece.neu.edu).
- S. Silvestri, and S.K. Das are with the Department of Computer Science, Missouri University of Science and Technology, Rolla, MO, 65401 USA (e-mail: {silvestris, sdas}@mst.edu).
- P. Ferraro and G. Lo Re are with the Department of Computer Engineering, University of Palermo, Palermo PA 90128 Italy (e-mail: {pierluca.ferraro, giuseppe.lore}@unipa.it).

sourced traffic updates, where Alice and Bob are the only participants (see Figure 1). In such system, the only mobility information available (obtained through the navigation app) is Alice and Bob's current location, their destination (L2), and their current route. We remark that in this scenario Alice and Bob's mobility is *uncertain*, since their arrival time and route may change significantly due to changing traffic conditions. Let us assume the current time is 10:00AM. Let us also assume the system needs two sensing tasks to be executed, namely T1 (traffic status in L1 at 10:00AM) and T2 (traffic status in L2 at 10:10AM). Given Alice and Bob's current geographic location and destinations, it would be reasonable to assign T1 to Alice and T2 to Bob. However, the system should also consider that it may not be possible for Bob to execute T2, due to a traffic jam along his route (see Figure 1), and thus assign T2 to both Alice *and* Bob.

**Challenges.** On the other hand, is it *feasible* from a budget perspective to assign T2 to Alice and Bob? In other words, is it possible to devise a mechanism able to schedule T2 to both Alice and Bob and also provide them with a satisfying reward, without exceeding a reward budget, by also considering uncertain mobility? More formally, we would like to achieve the following (very challenging) design objectives [10] when designing incentive mechanisms:

- The mechanism must assume that the budget for recruiting participants is often *limited*, thus not every participant can be hired by the system (i.e., the mechanisms must be *budget-feasible*);
- In order to prevent *churning*, the mechanism must be *truthful* and *individual-rational*, in the sense that it is the most rational choice for participants to keep participating because it is always convenient for them;
- Given the large scale of mobile crowdsensing systems, the mechanisms must be *computationally efficient*, which means the process of selecting and rewarding participants must have polynomial complexity and be possibly *parallelizable*.

**Novel Contributions.** The above mentioned reasons and challenges motivate our work and the following novel contributions, which are summarized as follows.

- We propose a novel framework named *IncentMe* to study and control the interaction between the mobile crowdsensing platform (MCP) and crowdsensing participants with uncertain mobility. These interactions are modeled as a *reverse auction* between a buyer/auctioneer (i.e., the MCP) and a group of sellers/bidders (i.e., the crowdsensing participants). Specifically, in our model, participants compete with each other for being selected to execute sensing tasks by submitting a bid containing their expected reward. The MCP uses the auction to select the winning bidders and compute the associated reward, in such a way that the sum of rewards does not exceed a budget and the likelihood of sensing task execution is maximized.
- To optimize the selection and reward algorithms of our reverse auction, we formulate the novel *Budgeted Value Maximization Problem* (BVM), and mathematically prove through reduction that BVM is an NP-Hard problem. To solve BVM, we leverage *game-theoretical mechanism design* [11] to derive two polynomial-time mechanisms, called by us *Truthful Value Maximization* (TVM) and *Heuristic Value Maximization* (HVM). We mathematically prove that TVM and HVM achieve an approximation ratio of $\approx 0.21$, which is to the best of our knowledge the greatest with respect to the state of the art in the field [12], [13]. We also mathematically prove that TVM and HVM guarantee the fundamental properties of truthfulness, individual-rationality, and budget-feasibility, and discuss how to parallelize TVM and HVM to achieve maximum performance.
- We experimentally evaluate the TVM and HVM auction mechanisms by considering a road traffic monitoring application where participants with uncertain mobility submit information about traffic events (e.g., accidents, traffic lines, etc.) during their trips. In order to realistically implement the application, and experiment with different mobility patterns, we use real-world mobility traces of taxi cabs in San Francisco [14], Rome [15], and Beijing [16]. We also evaluated the performance of HVM when multiple parallel jobs are used. Experimental results demonstrate that our mechanisms outperform the state of the art [12], [13], [17], [18] by improving on the average of 30% the likelihood of successful task execution, as well as achieving a significant speedup of about 12x in the scenarios considered.

**Paper Organization.** The paper is organized as follows. Section 2 introduces the system model and the problem definition, while Section 3 presents the budget-feasible mechanisms of IncentMe and Section 4 discusses the experimental results. Related work is summarized in Section 5, while Section 6 concludes the paper.

## 2 SYSTEM MODEL

In this section, we describe the architecture of IncentMe as well as mathematically formalize it. Hereafter, we will refer to the terms "user", "participant" and "bidder" interchangeably, as well as "system" and "MCP".

The proposed architecture is depicted in Figure 2, consisting of a platform (MCP) which can be accessed through wireless Internet connection. The interactions between the participants and the MCP are detailed below.

- Smartphone owners who intend to participate in the mobile crowdsensing campaign download the *mobile crowdsensing app* (step 1) through common markets such as *App Store* or *Google Play*. The app is responsible for handling data transmission, visualization, and bidding process.
- Whenever required, the MCP generates a set of *sensing tasks* that need to be executed (step 2). Sensing tasks specify a series of requirements, such as the sampling rate requested, minimum sensing time, maximum distance from specified location, or task expiration time.
- When users intend to participate, a *bid* must be submitted, which is the minimum amount of reward the

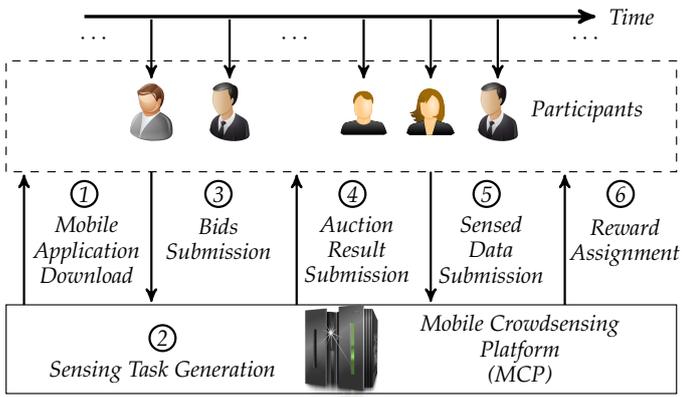

Fig. 2: Interactions between participants and MCP.

user would like to receive to execute a sensing task (step 3). The MCP may also collect mobility information about the user, which can be used to determine the user's availability to perform the sensing tasks.
- After receiving the bids, the MCP runs the auction, and based on its outcome, selects a subset of participants, assigns them the sensing tasks, and communicates the reward assigned to each user (step 4). After being selected and having received the list of sensing tasks to be executed, a participant is allowed to begin performing the sensing service (step 5). After the sensing tasks have been executed and the sensed data collected, the MCP rewards each user the amount in step 4 (step 6).

**Example of Use Case of IncentMe.** Consider a navigation app where the system crowdsources traffic-related info, similar to the example discussed in the Introduction. To participate, drivers download the app and register to the system, by providing (i) payment credential for depositing rewards, (ii) minimum reward for submitting each traffic report, and (iii) frequency of traffic report submission. Such info can be changed later by going to the Settings page of the app. Whenever a new trip begins, a driver enters the destination location by using the smartphone app (start location is automatically retrieved from GPS), and a bid is automatically submitted to the MCP as follows:

- Name: Alice Jones
- Unique ID: 1242760243
- Trip Start: 122 W 12th Street, Manhattan, NYC
- Trip End: 345 E 57th Street, Manhattan, NYC
- Reward: Minimum $0.1 for each traffic report
- Frequency: One traffic report every 10 minutes

Upon receiving such information, the MCP computes Alice's estimated time of arrival (ETA) and routing information by using GraphHopper or Google Maps APIs[1]. This information is then used to infer Alice's mobility distribution and run the auction mechanism (as explained in Section 3). This process will decide whether Alice will be selected and how much she will be rewarded. If selected, the MCP

---
1. Grasshopper available at https://graphhopper.com/ and Google Maps API at https://developers.google.com/maps/documentation/directions/intro

replies with the following message, which is displayed on the sensing app screen:

- Name: Alice Jones
- Unique ID: 1242760243
- Reward: $0.27 for each traffic report

Alice may then submit every 10 minutes a traffic report during her trip to 345 E 57th Street. Depending on the number of reports, the MCP then pays Alice accordingly. For example, if she submitted 6 reports, Alice will be paid $6 \times 0.27 = \$1.62$. Note here that Alice received a higher reward per report than what she asked for. This is possible, as explained in Section 3.

**Important Remarks.** We point out that the bidding process is handled automatically by the app, and is transparent to the participant (i.e., does not require any interaction). This aspect makes IncentMe easily embeddable in existing navigation apps.

TABLE 1: Summary of main symbols.

| Symbol | Description |
|---|---|
| $\mathcal{S}$ | set of sectors |
| $\mathcal{Q}$ | set of participants |
| $\mathcal{Z}$ | set of sensing tasks |
| $\mathcal{T}$ | set of winning bidders |
| $\mathcal{B}$ | set of bidders |
| $\mathcal{R}$ | vector of rewards |
| $\alpha$ | auction's allocation function |
| $\pi$ | auction's payment function |
| $B$ | auction's budget |
| $\tau_{i,j}$ | sensing task for sector $i$ at time $t_j$ |
| $V_{ij}$ | value of sensing task $\tau_{i,j}$ |
| $W(i,j,\mathcal{T})$ | prob. at least one participant is in sector $i$ at $t_j$ |
| $\mathcal{V}(\mathcal{T})$ | total value obtained by the auction |
| $p_k^{i,j}$ | mobility distribution of $k$-th participant |
| $\gamma_k$ | cost of $k$-th participant to execute a sensing task |
| $\nu_k$ | bid submitted by $k$-th participant |
| $u_k(\nu_k)$ | utility obtained by the $k$-th participant by bidding $\nu_k$ |

### 2.1 Definitions

Let us now formalize the IncentMe system described above. Table 1 summarizes the main symbols used throughout the paper. To model participants' mobility, the sensing area is divided into *sectors*, which represent the sensing granularity of the application. For example, in a traffic monitoring application, the sectors can be as large as a city block, whereas in air quality monitoring applications the sector can be as large as a neighborhood. Time is discretized, with $j$ being the $j$-th time slot between $t_j$ and $t_{j+T}$.

Let $\mathcal{S}$ define the set of $s = |\mathcal{S}|$ sectors forming the sensing area. Let $\mathcal{Q}$ be the set of $m = |\mathcal{Q}|$ participants competing for offering their sensing services. We define *sensing task* as a sensing activity that the MCP needs to be performed in a particular place and at a particular time. More formally:

**Definition 1.** *(Sensing task). A sensing task is a tuple $\tau_{i,j} = (i,j)$ where $i \in \mathcal{S}$ indicates the sector and $j \in \mathbb{R}^+$ indicates the timestep (e.g., $\tau_{3,4}$ indicates the sensing task involving sector 3 at timestep 4). We let $\mathcal{Z}$ indicate the set of sensing tasks.*

**Mobility Assumptions and Remarks.** We realistically assume that participants can leave and enter the system at their will. For simplicity, we define $t = 0$ as the moment in which the auction is executed. To model participants' mobility, let $p_k^{i,j}$ indicate the probability that the $k$-th participant will be in sector $i$ at time $j$. To derive proofs that are independent from a given participants' mobility pattern, our incentive mechanisms will be derived by considering generic $p_k^{i,j}$ values. As described in the use case in Section 3.1, participants' mobility information can be acquired (with some degree of uncertainty) by using APIs such as Google Maps or GrassHopper. Note that we are *not* assuming that participants' mobility is perfectly known. Furthermore, given the large number of participants to real-world mobile crowdsensing systems, we assume that the number of participants that could team up to manipulate the auction result is not significant w.r.t. the total number of participants; this assumption has also been made in existing related work [7].

We now formally define some terms that will be frequently utilized in this paper.

**Definition 2.** *(Mechanism) [11]. Let $\mathcal{B}$ be the set of bidders, $\mathcal{T}$ be the set of winning bidders, and $\mathcal{R}$ be the vector of rewards given to the auction participants. A mechanism $\mathcal{M}$ defines a tuple $\{\alpha, \pi\}$, where $\alpha : \mathcal{B} \to \mathcal{T}$ is defined as the auction's allocation function, and $\pi : \mathcal{T} \to \mathcal{R}$ is the auction's payment function.*

As regards to the bidding process, we know participants spend resources when performing sensing services, for example, personal time, device battery, and network bandwidth. We denote $\gamma_k$ as the cost of the $k$-th participant for executing a sensing task, which includes the minimum profit that the participant desires to earn by participating. We consider $\gamma_k$ as personal information and thus not revealed to the MCP. We also define $\nu_k$ as the bid submitted by participant $k$. We point out that in general, $\nu_k \neq \gamma_k$ (*i.e., the bid may be different from the cost*).

**Definition 3.** *(Utility) [11]. The utility obtained by the $k$-th auction participant by bidding $\nu_k$ and receiving as reward $\mathcal{R}(k)$ is the quantity $u_k(\nu_k) = \mathcal{R}(k) - \gamma_k$, where $\gamma_k$ is the cost incurred by $k$ when executing one sensing task.*

In the equation above, the reward $\mathcal{R}(k)$ assigned to each winning bidder $k \in \mathcal{T}$ depends on the bid $\nu_k$ through the auction's payment function $\pi : \mathcal{T} \to \mathcal{R}$. This implies that the utility $u_k$ also depends on $\nu_k$, hence its appearance on the left side of the equation.

Since participants are selfish and interested in maximizing their own utility, they may try to *overbid* (i.e., submitting $\nu_k$ much higher than $\gamma_k$), trying to achieve a higher reward. This may ultimately compromise the auction's efficiency [11]. To solve this problem, we need to design *truthful* mechanisms that align the participants' interests with the system goals. A mechanism is truthful if any user maximizes her utility by bidding her real cost $\gamma_k$, no matter how other participants may act; we also require our mechanisms to satisfy *individual rationality*, which means that any user always gets a non-negative utility, and to be computationally efficient. These properties are formally defined below.

**Definition 4.** *(Truthfulness) [11]. The mechanism $\mathcal{M} = \{\alpha, \pi\}$ is truthful iff $u_k(\gamma_k) \geq u_k(x)$, for any $x \neq \gamma_k$.*

**Definition 5.** *(Individual rationality) [11]. A mechanism is individual-rational iff $u_k(x) \geq 0, \forall k$.*

**Definition 6.** *(Computational efficiency) [11]. A mechanism is computationally efficient iff $\alpha$ and $\pi$ have at most polynomial complexity in the number of bids.*

Definition 6 is a standard definition: our mechanisms are indeed low-order polynomial and are scalable (i.e., they can be parallelized).

## 3 MECHANISM DESIGN

Figure 3 summarizes the flow of Section 3. First, the Budgeted Value Maximization (BVM) problem is introduced, then we show it is NP-Hard. Next, to provide mechanisms with approximate solutions, we demonstrate that BVM's value function is submodular. We then propose Truthful Value Maximization (TVM) mechanism to efficiently solve the BVM, and prove that it is truthful, individual-rational, budget-feasible, and achieves an approximation ratio with the optimum solution. To further improve over TVM, we propose Heuristic Value Maximization (HVM), a mechanism that leverages interpolation search [19] to provide a more efficient solution to the BVM.

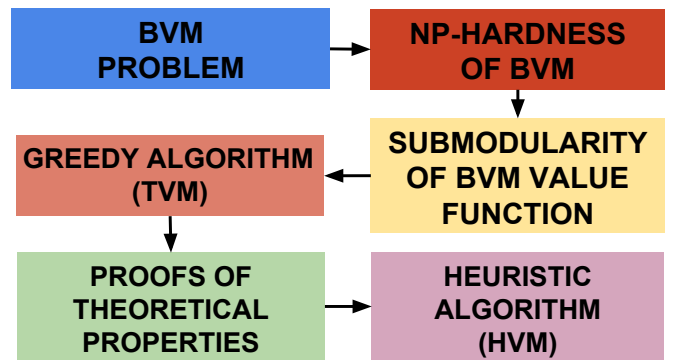

Fig. 3: Summary of Section 3.

### 3.1 Budgeted Value Maximization

We are now ready to formulate the user incentivization problem studied in this paper. Let us define $V_{ij}$ as the value that sensing task $\tau_{ij}$ has for the MCP. Intuitively, $V_{ij}$ models the preference that the system has for some sensing tasks instead of others (e.g., covering some neighborhoods of a city may be more important than covering others). For notational simplicity, $V_{ij} = 0$ if $\tau_{ij} \notin \mathcal{Z}$, the set of sensing tasks.

**Problem 1.** *Budgeted Value Maximization (in short, BVM). Let $W(i, j, \mathcal{T}) = 1 - \prod_{k \in \mathcal{T}}(1 - p_k^{i,j})$ be the probability that at least one participant will be in sector $i$ at time $j$, which is also the probability that at least one participant will be able to execute the sensing task. Finally, let us define $B$ as the available budget, $z$ as the maximum timestamp of the available sensing tasks, hereafter referred to as auction duration, and $s$ as the number of sectors. Given values $V_{ij}$ and function $W(i, j, \mathcal{T})$, let $\mathcal{V}(\mathcal{T})$ be defined as*

$$\mathcal{V}(\mathcal{T}) \triangleq \sum_{i=1}^{s} \sum_{j=1}^{z} V_{ij} \cdot W(i, j, \mathcal{T}) \tag{1}$$

*Find set of bidders $\mathcal{T}^*$ and reward vector $\mathcal{R}^*$ such as*

$$\mathcal{T}^* = \arg\max_{\mathcal{T} \subseteq \mathcal{B}} \sum_{i=1}^{s} \sum_{j=1}^{z} V_{ij} \cdot W(i, j, \mathcal{T}), \sum_{r \in \mathcal{R}^*} r \leq B$$

The optimization function defined in BVM can be seen as a function expressing the probability that the sensing tasks will be executed (each one weighted with its value).

Before introducing the novel mechanisms developed in this paper, we first demonstrate that BVM is NP-Hard.

**Theorem 1.** *BVM is NP-Hard.*

*Proof.* In order to prove the NP-Hardness of BVM, we provide a reduction from the well-known Bounded Knapsack problem (BKP). A general BKP instance has a capacity $B$ and a set of items $\Omega = \{s_1 \cdots s_n\}$, where each item $s_i \in \Omega$ has a value $v_i$ and a weight $w_i$. The goal of BKP is to find a set $S^* \subseteq \Omega$ whose items provide maximum value and do not exceed the capacity $B$ of the knapsack. We translate the general BKP instance into a simpler instance of BVM, where we consider a single time step (i.e., $z = 1$). We also assume that each bidder is in a different sector of the sensing area, the probability of being in that sector is equal to one, and that the payment function is such as $\rho_k = \nu_k$. By defining $X_k$ an indicator function that is equal to one if a bidder is in sector $k$, and zero otherwise, a solution to BVM can be translated into a solution of the BKP instance by setting the value $v_k$ of each knapsack item to $V_{k1} \cdot X_k$, its weight $w_k$ to $\nu_k$, and $B$ as the size of the knapsack. As a result, solving BVM is at least as hard as solving BKP, therefore BVM is NP-Hard. □

### 3.2 Submodularity of BVM

We now prove that the objective function of BVM (defined in Equation 1) belongs to the family of *submodular functions* [20]. Let us now define submodularity, and prove some properties necessary to design a greedy algorithm with proven approximation ratio.

**Definition 7.** *(Submodularity)* [20]. *Given ground set $S$ and a function $\mathcal{F} : 2^S \to \mathbb{R}^+$, then $\forall A \subseteq B \subseteq S$, $\mathcal{F}$ is submodular iff, for any $i \in S$, $\mathcal{F}(A \cup \{i\}) - \mathcal{F}(A) \geq \mathcal{F}(B \cup \{i\}) - \mathcal{F}(B)$.*

**Theorem 2.** *The function $\mathcal{V}(\mathcal{T})$, defined in Equation (1) is (i) submodular, (ii) non-decreasing, and (iii) $\mathcal{V}(\emptyset) = 0$.*

*Proof.* Let us define $\Delta_{X,k} \triangleq \mathcal{V}(X \cup \{k\}) - \mathcal{V}(X)$ as the increment in value to $\mathcal{V}$ given by the addition to $X$ of a generic element $k$, hereafter referred to as **marginal value** of $k$ given $X$. In order to prove (i), we need to show that, for sets $A \subseteq B \subseteq S$, it is true that $\Delta_{A,k} \geq \Delta_{B,k}$.

$$\Delta_{X,k} = \sum_{i=1}^{s}\sum_{j=1}^{z} V_{ij} \cdot \left[1 - \prod_{y \in X \cup \{k\}}(1-p_y^{i,j})\right] -$$
$$\sum_{i=1}^{s}\sum_{j=1}^{z} V_{ij} \cdot \left[1 - \prod_{y \in X}(1-p_y^{i,j})\right]$$
$$= \sum_{i=1}^{s}\sum_{j=1}^{z} V_{ij} \cdot \left[1 - (1-p_k^{i,j})\prod_{y \in X}(1-p_y^{i,j})\right] -$$
$$\sum_{i=1}^{s}\sum_{j=1}^{z} V_{ij} \cdot \left[1 - \prod_{y \in X}(1-p_y^{i,j})\right]$$
$$= \underbrace{\sum_{i=1}^{s}\sum_{j=1}^{z} V_{ij} \cdot p_k^{i,j} \cdot \prod_{y \in X}(1-p_y^{i,j})}$$

The braced section in $\Delta_{X,k}$ does not depend on $X$. Thus, it is only needed to prove that $\prod_{y \in A}(1-p_y^{i,j}) \geq \prod_{y \in B}(1-p_y^{i,j})$ holds for any $A \subseteq B$.

$$\underbrace{\prod_{y \in A}(1-p_y^{i,j})}_{\triangleq Z} \geq \prod_{y \in B}(1-p_y^{i,j}) \triangleq$$
$$Z \geq Z \cdot \prod_{y \in \{B-A\}}(1-p_y^{i,j}) \triangleq$$
$$1 \geq \prod_{y \in \{B-A\}}(1-p_y^{i,j})$$

Since $p_y^{i,j}$ is a probability, the inequality above holds. This proves property (i).

Furthermore, property (ii) can be derived straightforwardly from the fact that $W(i, j, \mathcal{T} \cup \{i\}) \geq W(i, j, \mathcal{T}) \; \forall \mathcal{T}$, and property (iii) follows from $W(i, j, \emptyset) = 0$, by definition of empty product. □

### 3.3 Truthful Value Maximization

Solving the BVM while guaranteeing truthfulness and budget-feasibility is extremely challenging. For this reason, we propose Truthful Value Maximization (TVM), which is a mechanism that adopts recent advances in the field of submodular function maximization [20] to provide a solution to the BVM with proven approximation ratio through a greedy strategy.

---
**Algorithm 1** TVM allocation function
---
**Input:** $B, \mathcal{B}, \mathcal{Q}, \mathcal{V}$
**Output:** $\mathcal{T}, \mathcal{T}_v$
1: $\mathcal{T} = \emptyset$
2: $\mathcal{T}_v = \emptyset$
3: $\mathcal{T}_c = \mathcal{Q}$
4: **while** $\mathcal{T}_c \neq \emptyset$ **do**
5: 
$$k^* = \arg\max_{k \in \mathcal{T}_c} \Delta_{\mathcal{T},k}/\nu_k$$
6:   **if** $\nu_{k^*} + \sum_{k \in \mathcal{T}} \nu_k \leq B$ **then**
7:     **if** $\nu_{k^*} \leq \frac{B}{2} \cdot \frac{\Delta_{\mathcal{T},k^*}}{\Delta_{\mathcal{T},k^*} + \sum_{v \in \mathcal{T}_v} \Delta_{\mathcal{T},v}}$ **then**
8:       Append $k^*$ to $\mathcal{T}$
9:       Append $\Delta_{\mathcal{T},k^*}$ to $\mathcal{T}_v$
10:     **end if**
11:   **end if**
12:   $\mathcal{T}_c = \mathcal{T}_c - \{k^*\}$
13: **end while**
14: **return** $\mathcal{T}, \mathcal{T}_v$
---

Algorithm 1 presents the allocation function of TVM. It incrementally constructs a set of winners $\mathcal{T}$, initially empty (line 1). At each iteration, the algorithm picks an unconsidered bidder $k^*$ having maximum weight, where the weight is defined as the increase in the function $\mathcal{V}$ that $k^*$ provides, divided by its bid (line 5). The bidder $k^*$ is included in $\mathcal{T}$ only if the current sum of bidding values is not exceeded (line 6) and if a condition regarding the new bid $\nu_k$ is satisfied (line 7). In Theorem 6, we demonstrate that

**Algorithm 2** TVM payment function

**Input:** $\mathcal{T}, B, \mathcal{B}, \mathcal{Q}, \mathcal{V}$
**Output:** $\mathcal{R}$
1: **for** every $i \in \mathcal{T}$ **do**
2: $\quad \mathcal{B}^i = \mathcal{B} - \{\nu_i\}, \mathcal{Q}^i = \mathcal{Q} - \{i\}$
3: $\quad \mathcal{T}^i, \mathcal{T}_v^i$ = Algorithm-1($B, \mathcal{B}_i, \mathcal{Q}_i, \mathcal{V}$)
4: $\quad \mathcal{X} = \emptyset$
5: $\quad$ **for** every $j = 1 \ldots |\mathcal{T}^i|$ **do**
6: $\quad\quad \nu_{ij} = \frac{\Delta_{\mathcal{X},i} \times \nu_j}{\Delta_{\mathcal{X},j}}$
7: $\quad\quad \rho_{ij} = \frac{B}{2} \cdot \frac{\Delta_{\mathcal{X},i}}{\sum_{j' \leq j} \mathcal{T}_v^i(j) + \Delta_{\mathcal{X},i}}$
8: $\quad\quad$ Append $\mathcal{T}^i(j)$ to $\mathcal{X}$
9: $\quad$ **end for**
10: **end for**
11: **return** $\mathcal{R}(i) = \max_j \{\min\{\rho_{ij}, \nu_{ij}\}\}, \forall i \in \mathcal{T}$

the stopping criterion in line 7 ensures budget-feasibility of the mechanism and a constant approximation ratio with respect to the optimum solution of BVM. The algorithm returns the set $\mathcal{T}$ of winning bidders.

The payment scheme of TVM, as described in Algorithm 2, assigns to each winning bidder a payment corresponding to the *critical value*, which provably ensures truthfulness of the mechanism [10]. We point out that the critical value of a bidder is the maximum value that bidder could have bid and still win the auction. Unfortunately, the critical value computation is complicated by the submodularity of the marginal contributions $\Delta_{\mathcal{T},k}$, which implies that the value depends on the order by which participants are selected in the allocation function of TVM. To compute payments in an efficient way, for each bidder $i$, we consider the maximum of all possible bids that $i$ could have declared and still get allocated, as explained next. Consider running the allocation function without $i$. Let us consider the first $j$ participants with highest marginal value. By using the marginal contribution of $i$ at point $j$, we can find the maximal cost that agent $i$ can declare in order to be allocated instead of the agent in the $j$-th place in the sorting. Therefore, paying the user the maximum over all the possible $j$ positions ensures that $i$ is paid the critical value.

**Example.** For simplicity, we consider a sensing area composed of four sectors and a time range of only 1 timestep, so $z = 1$. Let us consider the case in which three bidders are competing to offer their sensing services. The value of each sector is $V = \{.3, .2, .1, .4\}$, while the mobility distributions of the bidders at timestep 1 are as follows: $p_1^1 = \{.2, .1, .3, .4\}$, $p_2^1 = \{0, .8, .05, .15\}$, $p_3^1 = \{.4, .2, 0, .4\}$. The bids submitted are $\mathcal{B} = \{10, 8, 12\}$.

*Allocation function*

Input: $B = 20, \mathcal{B} = \{10, 8, 12\}$

- **Step 1:** $\mathcal{T} = \{\}, \mathcal{T}_v = \{\}$

  $\frac{\Delta_{\mathcal{T},1}}{\nu_1} = \frac{0.2 \cdot 0.3 + 0.1 \cdot 0.2 + 0.3 \cdot 0.1 + 0.4 \cdot 0.4}{10} = 0.027$
  $\frac{\Delta_{\mathcal{T},2}}{\nu_2} = \frac{0 \cdot 0.3 + 0.8 \cdot 0.2 + 0.05 \cdot 0.1 + 0.15 \cdot 0.4}{8} = 0.028$
  $\frac{\Delta_{\mathcal{T},3}}{\nu_3} = \frac{0.4 \cdot 0.3 + 0.2 \cdot 0.2 + 0 \cdot 0.1 + 0.4 \cdot 0.4}{12} = 0.0277$
  Is $8 \leq \frac{20}{2} \cdot \frac{0.28125}{0.28125}$ ? YES

  Append $k^* = 2$ to $\mathcal{T}$
  Append $\Delta_{\mathcal{T},k^*} = 0.225$ to $\mathcal{T}_v$

- **Step 2:** $\mathcal{T} = \{2\}, \mathcal{T}_v = \{0.225\}$

  $\frac{\Delta_{\mathcal{T},1}}{\nu_1} = \frac{0.2285}{10} = 0.02285$
  $\frac{\Delta_{\mathcal{T},3}}{\nu_3} = \frac{0.2640}{12} = 0.022$
  Is $10 \leq \frac{20}{2} \cdot \frac{0.02285}{0.02285 + 0.225}$ ? NO

  Condition on budget (line 7) is not fulfilled. Algorithm 1 terminates, and returns $\mathcal{T} = \{2\}$ and $\mathcal{T}_v = \{0.225\}$

*Payment function*

Input: $\mathcal{T} = \{2\}$

- **Step 1:** $i = 2$
  Run auction without 2 (lines 2-3)

  – **Step 1-A:** $j = 1$

    $\nu_{21} = 0.225 \cdot {}^{10}/_{0.27} = 8.33$
    $\rho_{21} = 10$

  To win against 1, user 2 has to bid 8.33.

Return $\mathcal{R}(2) = \max_{j \in \{1\}} \{\min\{\rho_{2j}, \nu_{2j}\}\} = 8.33$

### 3.4 Proof of Theoretical Properties

We now prove that TVM is truthful, individual-rational, and budget-feasible.

**Theorem 3.** *TVM is truthful.*

*Proof.* In order to characterize the truthfulness of the mechanism, we apply the characterization of Myerson [10]. A mechanism is truthful iff (i) the allocation function is monotone: if bidder $k$ wins the auction by bidding $\nu_k$, it also wins by bidding $\nu_k' < \nu_k$; (ii) Each winner is paid the critical value: bidder $k$ would not win the auction if it bids higher than the critical value.

(*Monotonicity*). The first property is guaranteed by the greediness of the algorithm. By lowering her bid, any allocated bidder would only increase her marginal value per unit cost and thus be placed ahead in the sorting order considered by the allocation function.

(*Critical value*). According to Algorithm 2, each winning bidder $i$ is rewarded $\max_j \{\min\{\rho_{ij}, \nu_{ij}\}\}$. Let us consider $r$ to be the index for which $P_i = \min\{\rho_{ir}, \nu_{ir}\}$. Therefore, bidding $P_i$ implies that $i$ would be allocated at position $r$ in the run of the algorithm without $i$. Four different cases are thus possible.

1) $\nu_{ir} \leq \rho_{ir}$ and $\nu_{ir} = \max_j \nu_{ij}$. Reporting a bid higher than $\nu_{ir}$ places bidder $i$ after the first unallocated user $k^*$ in the alternate run of the mechanism, thus $i$ would not be selected.
2) $\nu_{ir} \leq \rho_{ir}$ and $\nu_{ir} < \max_j \nu_{ij}$. Consider some $j$ for which $\nu_{ir} < \nu_{ij}$. Since $r$ has maximality condition, it must be the case that $\rho_{ij} \leq \nu_{ir} \leq \nu_{ij}$. Therefore, bidding higher than $\nu_{ir}$ would violate the selection condition (line 7, Algorithm 1) and hence $i$ would not be allocated. For some other $j$ such as $\nu_{ir} \geq \nu_{ij}$, bidding higher than $\nu_{ir}$ would place $i$ after $j$, so $i$ would not be allocated at position $j$.

3) $\nu_{ir} \geq \rho_{ir}$ and $\rho_{ir} = \max_j \nu_{ij}$. Reporting a bid higher than $\rho_{ir}$ violates the selection condition at each of the indices in $j \in \mathcal{T}^i$, hence $i$ would not be selected.

4) $\nu_{ir} \geq \rho_{ir}$ and $\rho_{ir} < \max_j \nu_{ij}$. Consider some $j$ for which $\rho_{ir} < \rho_{ij}$. Since $r$ has maximality condition, it must be the case that $\nu_{ij} \leq \rho_{ir} \leq \rho_{ij}$. Therefore, bidding higher than $\rho_{ir}$ would put $i$ after $j$ and hence $i$ would not be allocated. For $j$ such as $\rho_{ir} \geq \rho_{ij}$, bidding higher than $\nu_{ir}$ would mean $i$ would not be allocated at considered position $j$.

In all four cases, bidding higher than $P_i$ would cause bidder $i$ to not be selected, thus $P_i$ is the critical value. □

**Theorem 4.** *TVM is individual-rational.*

*Proof.* Consider the bid that $i$ can declare to be allocated at position $j = i$ (i.e., back at its original position) in the alternate run of the mechanism. Therefore, the payment that $i$ will receive will be $P_i = \min\{\rho_{ii}, \nu_{ii}\}$, We prove that $\nu_i \leq P_i$. First, we show that $\nu_{ii} \geq \nu_i$:

$$\nu_{ii} = \frac{\Delta_{\mathcal{X},i} \cdot \nu_i}{\Delta_{\mathcal{X},j}} \geq \frac{\Delta_{\mathcal{X},i} \cdot \nu_i}{\Delta_{\mathcal{X},i}} = \nu_i$$

The equality holds because $\nu_j/\Delta_{\mathcal{T},j} \geq \nu_i/\Delta_{\mathcal{T},i}$ since $i$ was selected after $i-1$ in the selection algorithm instead of $j$. Next, we show that $\rho_{ii} \geq \nu_i$:

$$\rho_{ij} = \frac{B}{2} \cdot \frac{\Delta_{\mathcal{X},i}}{\sum_{j' \leq i-1} \mathcal{T}_v^i(j') + \Delta_{\mathcal{X},i}}$$
$$= \frac{B}{2} \cdot \frac{\Delta_{\mathcal{X},i}}{\sum_{j \leq i-1} \Delta_{\mathcal{X},j} + \Delta_{\mathcal{X},i}} \geq \nu_i \quad (2)$$

The second equality holds from the fact that the first $i-1$ allocated elements in both the runs of the mechanism are the same. The third inequality follows from the proportional share criteria used to decide the allocation of $i$ after $i-1$ participants were selected already. □

We provide sketches of Theorems 5 and 6 due to space limitations.

**Theorem 5.** *TVM is budget-feasible.*

*Proof.* (Sketch.) It is first proven that the maximum payment for a user $p$ is $2 \cdot \Delta_p / \sum_{k \in \mathcal{T}} \Delta_k \cdot B$, where $\Delta_p$ is the marginal contribution given by $p$ computed during the run of the allocation function. Thus, $\sum_{i \in \mathcal{T}} \mathcal{R}(i) \leq \sum_{i \in \mathcal{T}} 2 \cdot \Delta_p / \sum_{k \in \mathcal{T}} \Delta_k \cdot B/2 \leq B$. □

An approximation algorithm $A$ is said to have an approximation ratio (or factor) $\rho$ if for each input $x$ the value $f(x)$ of the approximate solution $A(x)$ will not be less than $\rho$ times the optimal value OPT. More formally, for algorithm $A$ it holds that

$$\rho \cdot \text{OPT} \leq f(x) \leq \text{OPT} \quad \forall x. \quad (3)$$

In other words, the approximation ratio is a measure of the performance of an approximation algorithm, since it shows the worst-case performance of the algorithm.

**Theorem 6.** *TVM achieves an approximation ratio of $(e - 1/3e) - \lambda \sim 0.2107 - \lambda$, where $\lambda$ is the ratio of the participants' largest marginal contribution and the optimum.*

*Proof.* (Sketch.) It is proven that TVM achieves a utility that is at least $1/3$ of the well-known greedy algorithm proposed in [21], which in turn achieves an approximation ratio of $e - 1/e - \lambda$. □

For example, if each participant can contribute at most 1% to the optimal utility (i.e., $\lambda = 0.01$), Theorem 6 guarantees a constant approximation factor of 0.2007.

**Theorem 7.** *Algorithm 1 has complexity $\Theta(m^2 \cdot s \cdot z)$, where $m$ is the number of bidders, $s$ is the number of sectors, and $z$ is the auction allocation span, while Algorithm 2 has complexity $\mathcal{O}(m^3 \cdot s \cdot z)$.*

*Proof.* The complexity of Algorithm 1 is dominated by the complexity of the while loop (line 4 through 15). At each iteration, the loop computes the quantities $\Delta_{\mathcal{T},k}$ for each $k \in \mathcal{T}_c$. As every iteration of loop removes one element from $\mathcal{T}_c$, this implies that this computation is performed $m + m - 1 + m - 2 + \cdots + 1 = \Theta(m^2)$ times. The complexity of computing $\Delta_{\mathcal{T},k}$ for a generic $k$ is dominated by the computation of

$$\mathcal{V}(\mathcal{T} \cup \{k\}) = \sum_{i=1}^{s} \sum_{j=1}^{z} V_{ij} \cdot W(i, j, \mathcal{T} \cup \{k\}) \quad (4)$$

We now provide a way to recursively derive in constant time $W(i, j, \mathcal{T} \cup \{k\})$ as a function of $W(i, j, \mathcal{T})$. The overall complexity of computing $\mathcal{V}(\mathcal{T} \cup \{k\})$ will be therefore $\Theta(s \cdot z)$, which yields a total algorithm complexity of $\Theta(m^2 \cdot s \cdot z)$.

$$\begin{aligned} W(i, j, \mathcal{T} \cup \{k\}) &= 1 - \prod_{q \in \mathcal{T} \cup \{k\}} (1 - p_q^{i,j}) \\ &= 1 - (1 - p_k^{i,j}) \cdot \overbrace{\prod_{q \in \mathcal{T}} (1 - p_q^{i,j})}^{1 - W(i,j,\mathcal{T})} \\ &= 1 - (1 - p_k^{i,j}) \cdot (1 - W(i, j, \mathcal{T})) \end{aligned} \quad (5)$$

where $W(i, j, \emptyset) = 0$ by definition. Since the algorithm might terminate before having evaluated every bidder (line 7), the complexity becomes $\mathcal{O}(m^2 \cdot s \cdot z)$. The complexity of Algorithm 2 can be calculated by observing that there are at most $m$ iterations of the main loop (line 1), and in each loop the complexity is dominated by the execution of the selection function (line 3). This yields a total complexity of $\mathcal{O}(m^3 \cdot s \cdot z)$. □

### 3.5 Heuristic Value Maximization

Although TVM achieves a provable approximation bound, the stopping criterion of TVM allocation algorithm (line 7, Algorithm 1) limits the efficiency of TVM, as a significant part of the budget will be left unutilized during the run of the payment function. In order to optimize TVM without sacrificing to efficiency, the monotonicity of the sum of payments can be leveraged. We now illustrate this point through an example.

**Example.** In the following, we will refer as *actual budget* ($B$) the budget available for the current auction, and by *input budget* ($\hat{B}$) the budget that is given as input to the TVM selection algorithm (line 1, Algorithm 1). Figure 4 depicts the sum of payments assigned to winning bidders by TVM as a function of the input budget value $\hat{B}$, where we consider 1000 bidders and truncated normal bid distribution (with mean=0.5 and support=1). As we can see from this plot, the sum of payments allocated to the winning bidders remains significantly below the actual budget. For example, assuming actual budget $B$ equal to 10, the optimal input budget value $B^*$ that yields sum of payments equal to 10 would be 31. Furthermore, $\hat{B}$ equal to 10 yields sum of payments equal to 2, which means that 80% of the allocated budget remains not utilized by TVM.

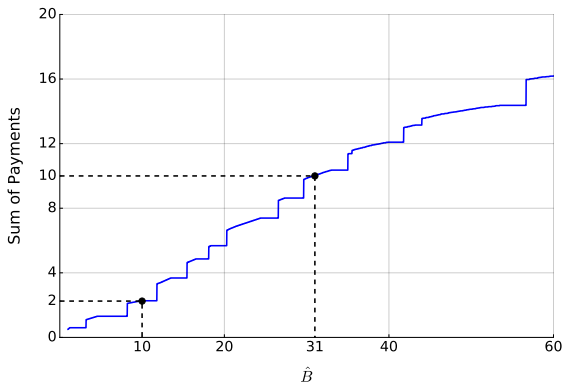

Fig. 4: TVM: sum of payments vs. input budget.

From this example, it emerges that the performance of TVM can be further optimized by exploring the input budget space and finding the $\hat{B}$ value yielding the highest sum of payments that remains below the actual budget. To explain our approach, let us assume to have an "array" of $n$ elements in which we store the sum of payments corresponding to all the input budgets from $B$ to $B+n-1$. Since payments are monotone with the budget, the array will be sorted. Thus, finding the optimum input budget corresponds to finding the rightmost place where the given sum of payments can be correctly inserted in the array without compromising the sorted order. However, given the values of the "array" are unknown in advance, we want to reduce as much as possible the number of guesses, since each guess implies running the TVM auction with given input budget.

Algorithm 3 introduces *Heuristic Value Maximization* (HVM), which uses interpolation and exponential searches [19] to find the optimum input budget by computing no more than $\mathcal{O}(\log \log n)^2$ times the TVM auction. HVM is mainly made up by two procedures:

- First, it finds the lower and upper bound $B_{min}$ and $B_{max}$ of the search interval by using exponential search, which doubles the search interval until the sum of payments is below the actual budget (line 2). This guarantees that the search interval will be found in $\mathcal{O}(\log B^*)$ runs of TVM auction.
- Then, it calculates the next point $B_{cur}$ by calculating the line passing between the two points

2. Bound valid under the assumption of uniformly distributed sum of payments [19].

**Algorithm 3** Heuristic Value Maximization
**Input:** $B, \mathcal{B}, \mathcal{Q}, \mathcal{V}$
**Output:** $B^*$
1: $B_{min} = B, B_{max} = B+1$
2: $B_{min}, B_{max}, \mathcal{R}_{min}, \mathcal{R}_{max} = \text{FMMB}(B, \mathcal{B}, \mathcal{Q}, \mathcal{V})$
3: **while** $B_{min} \leq B_{max}$ **do**
4:     $B_{cur} = \text{INT}(B_{min}, B_{max}, \mathcal{R}_{min}, \mathcal{R}_{max})$
5:     $\mathcal{T}_{cur} = \text{Algorithm-1}(B_{cur}, \mathcal{B}, \mathcal{Q}, \mathcal{V})$
6:     $\mathcal{R}_{cur} = \text{Algorithm-2}(\mathcal{T}, B_{cur}, \mathcal{B}, \mathcal{Q}, \mathcal{V})$
7:     **if** $\sum_{k \in \mathcal{T}_{cur}} \mathcal{R}_{cur}(k) > B$ **then**
8:         $B_{max} = B_{cur} - 1$
9:         $\mathcal{R}_{max} = \mathcal{R}_{cur}$
10:     **else**
11:         $B_{min} = B_{cur} + 1$
12:         $\mathcal{R}_{min} = \mathcal{R}_{cur}$
13:     **end if**
14: **end while**
15: Return $B_{min}$

$(B_{min}, \mathcal{R}_{min})$ and $(B_{max}, \mathcal{R}_{max})$ and computing the next step $B_{cur}$ as the $x$ component of the point passing by the actual budget $B$. If the payment $\mathcal{R}_{cur}$ yielded by the new point $B_{cur}$ is greater than (resp. less than or equal to) the budget, the algorithm explores the left (resp. right) part of the search interval, which is $[B_{min}, B_{cur} - 1]$ (resp. $[B_{cur} + 1, B_{max}]$) until the exit condition is met (line 4). In order to further speed up the execution time, the algorithm approximates the $\mathcal{R}_{max}$ and $\mathcal{R}_{min}$ values with the $\mathcal{R}_{cur}$ value (lines 9 and 12).

**Example.** Figure 5 shows a run of the exponential search used to find the search interval of HVM. In this example, the actual budget is $B = 15$, whereas the optimum is the input budget $B^* = 114$, corresponding to a sum of payment $P^* = 15$. By starting from $\hat{B}_0 = 15$ (same as actual budget), the interval is doubled until the sum of payment is greater than the actual budget, which is $\hat{B}_3 = 120$. The lower bound of the search interval will be $\hat{B}_2 = 60$. Thus, the starting interval is $[\hat{B}_2, \hat{B}_3] = [60, 120]$.

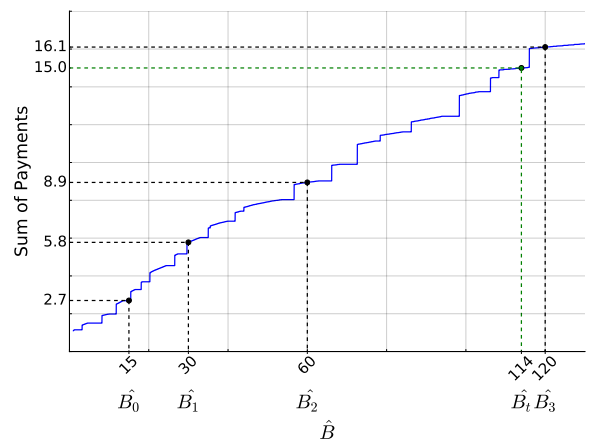

Fig. 5: Example of Exponential search.

Figure 6 shows the execution of one step of interpolation search; we also consider one step performed by binary search for comparison. The next point selected by binary search will be $\hat{B}_{cur} = \hat{B}_2 + \hat{B}_3 - \hat{B}_2/2 = 90$, which is the mid point of the interval. As far as interpolation search is concerned, the goal is to compute the intersection between the segment $(\hat{B}_2, \hat{P}_2) - (\hat{B}_3, \hat{P}_3)$ with the line $B = 15$. To this end, the slope of the segment is computed as $\Delta = (\hat{B}_3 - \hat{B}_2)/(\hat{P}_3 - \hat{P}_2) = 8.37$. Then, the next point is computed as $\hat{B}_{cur} = \hat{B}_2 + B - \hat{P}_2/\Delta = 60 + (15 - 8.945) \cdot 8.37 = 110$. As we can see, the computation of the next point provided by interpolation search is closer to the optimum value than the one provided by binary search.

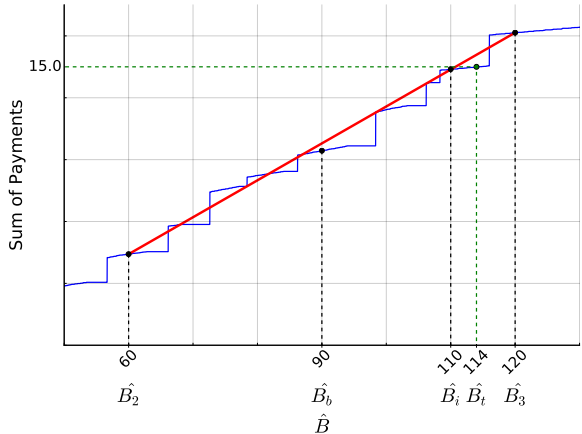

Fig. 6: Example of Interpolation search.

To conclude this section, we point out that the bounds and theoretical properties of HVM are the same to those of TVM, since HVM uses the same allocation and payment functions of TVM. Therefore, the result provided by HVM cannot be worse than TVM. We omit the formal proofs due to space limitations.

### 3.6 Parallel Implementation

Given the large number of participants in mobile crowdsensing applications, it is imperative to design incentive mechanisms that are able to handle the bidding of participants at scale. Thus, the objective is to adapt Algorithms 1 and 2 (i.e., selection and payment functions) to a parallel version, to handle a large number of bids from the participants. To this end, we notice that the main bottleneck of Algorithm 1 is line 5, which is the computation of the maximum value of $\Delta_{\mathcal{T},k}$ values, for every participant $k$. However, each of these quantities can be computed separately by different jobs, and the result reduced to obtain the maximum. Regarding the payment function (Algorithm 2), we notice that the computation of the payment for each winning bidder (line 1) can be assigned to a different job. The same can be applied to compute line 5. In Section 4, we show the performance evaluation of HVM when implemented with parallel jobs.

## 4 PERFORMANCE EVALUATION

In this section, we present the results obtained by evaluating the performance of HVM and TVM and the comparison of their performance with existing relevant work.

### 4.1 Experimental Setup

In order to obtain real-world participants' mobility and evaluate the performance with different mobility patterns, we have considered real mobility traces collected from the following datasets:

- *CRAWDAD-SanFrancisco* [14]: This dataset contains mobility traces of approximately 500 taxis in San Francisco, USA, collected over one month's time;
- *CRAWDAD-Rome* [15]: In this dataset, 320 taxi drivers in the center of Rome were monitored during March 2014;
- *MSR-Beijing* [16]: This dataset collected by Microsoft Research Asia contains the GPS positions of 10,357 taxis in Beijing during one month.

The heatmaps of the three datasets are shown in Figure 7, where warmer colors indicate the most popular places. As we can see, the mobility is more concentrated in Rome and Beijing, whereas it is more uniform in San Francisco.

The evaluation of the proposed mechanisms has been performed by emulating an application where taxi cab drivers report vehicular traffic events; its use case has been discussed in details in Section 2.1. In details, we consider the sensing area to be approximately 4×4km square areas, a timestep of 5 minutes, and the sensing areas divided into 400 sectors, such that each sector is approximately as large as a city block and the timestep is coherent to the granularity required by traffic monitoring applications.

To emulate bidders' behavior, inspired by existing work [22], we assume that (i) the duration of each taxi trip follows a Poisson probability distribution having average $\lambda_d = 6$ timesteps; and (ii) the time between each bid (*i.e.*, the bidding frequency) follows a Poisson probability distribution having average $\lambda_b = 3$ timesteps. The trip start and end locations, as well as the path followed by the taxis, are taken from the mobility traces. The MCP generates the sensing tasks in such a way that each sector of the sensing area must be covered during each timestep.

For the distribution of the bidding quantities, we were also inspired by previous work [22] and considered Gaussian distributed costs with mean 0.5 and standard deviation of 0.15. In all experiments, we considered 100 bidders, if not stated otherwise. In each experiment, we performed 100 repetitions and computed 95% confidence intervals, which are not shown if below 1%. As far as the weight of each sector, we have obtained the weight values by computing the sectors that are were most popular from the traces. For example, assuming that the sensing area had three sectors, if they were visited 10%, 50%, and 40%, respectively, the weights would have been set to $0.1, 0.5$, and $0.4$.

The experiments were performed on a cloud computing system emulated by two Dell Precision T7610 servers, equipped with an Intel(R) Xeon(R) CPU E5-2680 v2 processor (20 cores, 2.80GHz, 64GB RAM), and by four Dell Optiplex(R) 7010 with Intel(R) Core(TM) i7-3770 CPU (7 cores, 3.40GHz, 8GB RAM).

### 4.2 Comparison with Existing Work

For comparison reasons, we implemented the auctions due to Singer (IEEE FOCS 2010 [12]), Chen *et al.* (ACM SODA

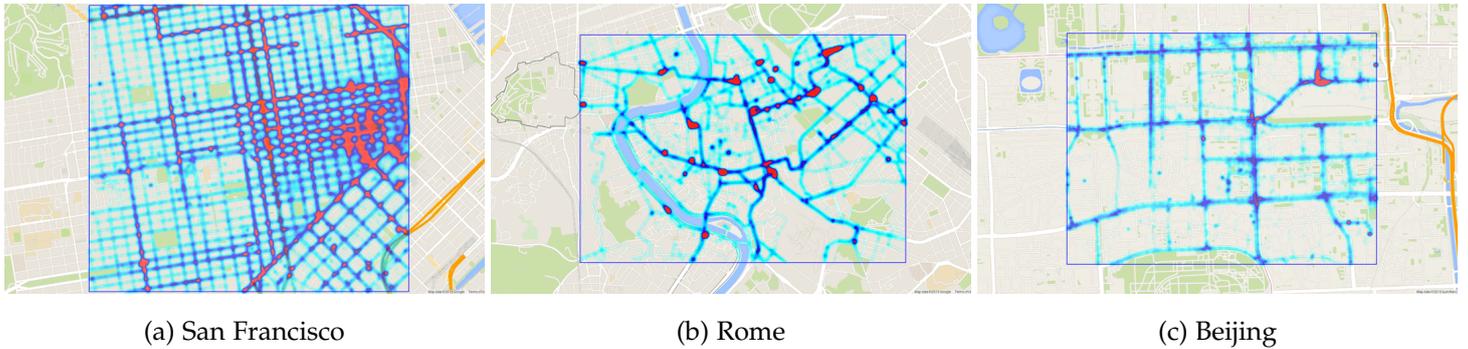

(a) San Francisco  (b) Rome  (c) Beijing

Fig. 7: Heatmaps of the mobility traces contained in the datasets.

2011 [13]). To the best of our knowledge, Singer and Chen are considered the most relevant work in the field of truthful incentive mechanisms with budget constraints. To compare HVM and TVM with budget-unconstrained approaches, we also implemented the IMCU auction (IEEE/ACM TNET 2016 [18]) and and the auction proposed by Hu *et al.* (IEEE/ACM IWQoS 2016 [17]), which is also tailored to address incentivization in vehicle-based mobile crowdsensing.

Similarly to us, Singer and Chen also use the theory of budgeted maximization of submodular functions, by providing randomized auction mechanisms with proven expected approximation ratio. However, as seen in Table 2, TVM and HVM outperform them by achieving an approximation factor that is 1.67x and 23.51x higher than Chen and Singer, respectively. This is because we do not use a randomized approach but instead we select participants based on their marginal values and make sure the total payment will be below the budget. On the other hand, Yang and Hu take a different approach and use the bid values as stopping condition of the selection function. Although this makes the algorithms simpler, Yang and Hu suffer of decreased performance, as seen in Section 4.3. Also, no maximum budget and no approximation bound is provided.

TABLE 2: Comparison of Approximation factors, TVM/HVM vs. Prior Work.

| Auction | Approximation Factor |
|---|---|
| Singer [12] | $\frac{(e-1)^2}{(58e^2-32e+6)} \approx 0.008929$ |
| Chen [13] | $\frac{5 \cdot e}{e-1} \approx 0.1264$ |
| Our work (HVM/TVM) | $\frac{3 \cdot e}{e-1} \approx 0.2107$ |

### 4.3 Experimental Results

In the following, we will use the *Obtained Value* (OV) and the *percentage of obtained value (POV)* to evaluate the performance of the auctions. The POV is defined as the success probability of the weighted sensing task obtained by the auction mechanism (see BVM problem, Equation 1) divided by the maximum obtainable value given by the optimum solution. More formally, if $P^*$ is the optimum, then $POV = {OV}/{P^*}$. Intuitively, the POV describes how efficient the mechanism is in recruiting participants, while the OV express the performance of each algorithm as absolute value. We will also use the *speed-up* to evaluate the scalability of HVM. Specifically, given execution time $E_1$ and $E_2$, the speed-up is defined as ${E_1}/{E_2}$. Since in this paper we are only considering the problem of effective and efficient recruiting of participants in mobile crowdsensing, we believe these metrics are sufficient and appropriate to evaluate and compare the performance of the mechanisms.

**Efficiency of TVM and HVM.** Figures 8 and 9 show the average POV and OV experimented by the considered mechanisms, as a function of the allocated budget per timestep. From the plots we can derive the following conclusions. First, HVM performs better than the other auctions, since HVM is significantly budget-effective as it uses almost all the available budget at each execution of the algorithm, while at the same time guaranteeing budget-feasibility. Second, TVM increases its performance significantly as the available budget increases, which is expected as the auction will terminate after selecting more participants. Third, the performance of existing auctions remains always below 70%, irrespective of the available budget. Moreover, the performance of Yang and Hu remains almost constant as it considers only the participants' bids and not the budget in the selection process. On the average, HVM increases the POV by a remarkable 33.2%, which is average of 31.5%, 36.3.%, and 31.9%. We also observe that the auctions obtain higher POV when the mobility is less dispersed (Rome and Beijing). Intuitively, this is because the sensing tasks that can be potentially executed are less, therefore it is more likely that participants will be able to execute them.

**Impact of Participants' Mobility on Performance.** As anticipated earlier, factoring participants' mobility into the user selection process is fundamental to obtain high performance of the incentive mechanism. To demonstrate this point, we show the POV and OV as a function of the probability that a bidder will not be able to execute the sensing task after been assigned to it, defined for simplicity as *task failure probability* (TFP). Note that this probability models effectively the case of uncertain mobility, as the effect of not being at a particular sector at a given moment in time (because the mobility has changed) implies that the assigned task will fail to be executed. As expected, Figures 10 and 11 show that the overall performance of all the algorithms decreases as the TFP increases. On the other hand, we point out that HVM is more resilient than the other algorithms, as the increased efficiency in budget utilization allows to select more bidders and in turn allows more redundancy in the selection process (i.e., hire more participants). On the average, HVM achieves 25.6% more POV (average of 24.8%, 27.5%, and 24.6%), which makes it ideal in cases when the TFP is high as it utilizes the allocated budget efficiently.

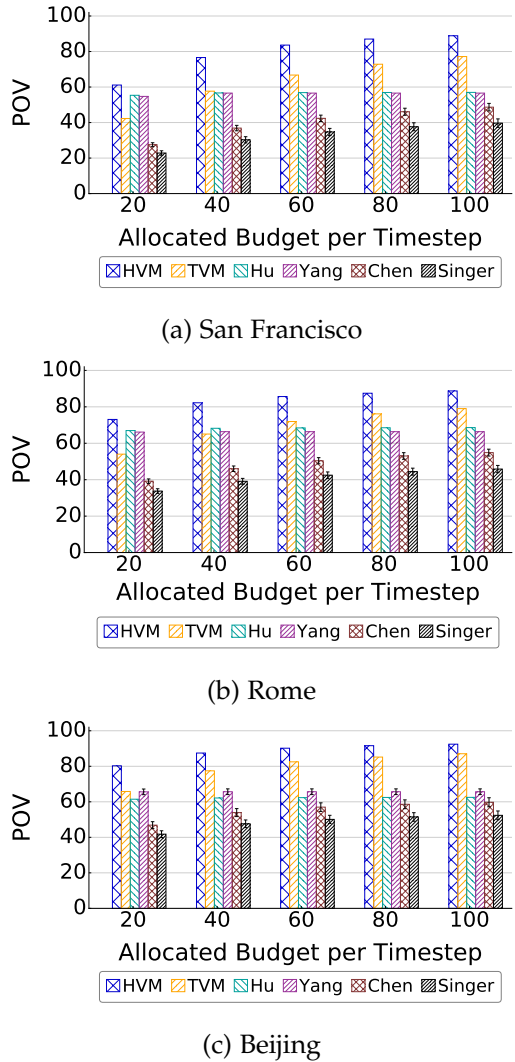

Fig. 8: POV vs. Allocated Budget per Timestep.

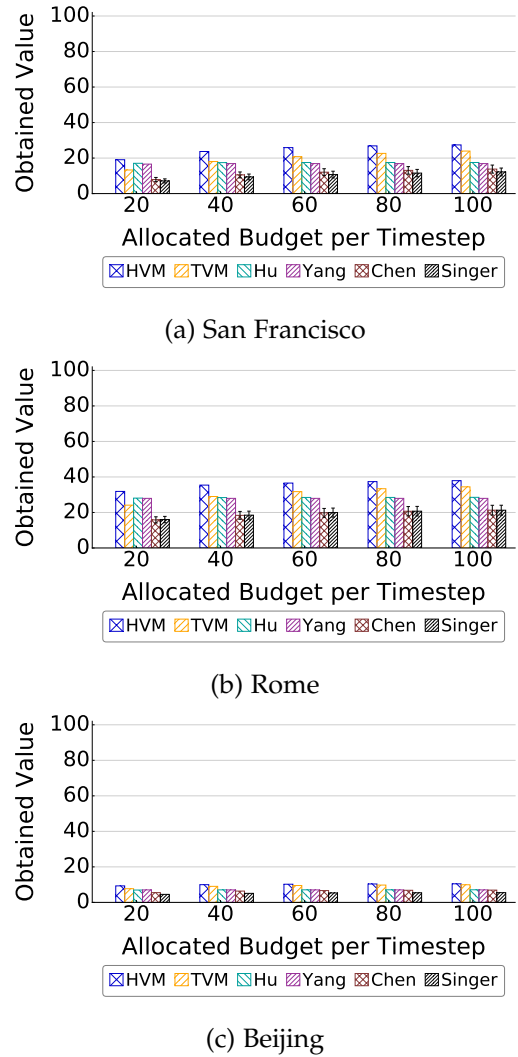

Fig. 9: Obtained Value vs. Allocated Budget per Timestep.

TABLE 3: Improvement of HVM vs Binary Search.

| Bidders # | Binary Search | HVM |
|---|---|---|
| 100 | 1.000 | 0.598 |
| 200 | 3.356 | 1.923 |
| 300 | 6.602 | 3.629 |
| 400 | 9.884 | 5.677 |
| 500 | 13.898 | 7.801 |
| 600 | 17.561 | 9.82 |
| 700 | 21.791 | 12.227 |
| 800 | 26.828 | 16.000 |
| 900 | 33.631 | 18.934 |
| 1000 | 36.660 | 21.012 |

**Impact of Parallelization on Performance.** Scalability is a fundamental asset of incentive mechanisms designed for mobile crowdsourcing contests. To this end, in order to demonstrate the scalability of HVM, Figure 12 shows the computation time as a function of the number of bidders and the number of jobs used in the parallel version of HVM. For clarity, we normalized the execution times by the same factor, so as to have one computation unit in the case the number of bidders is 200 and only 1 job is used. Figure 12 concludes that the speedup provided by the additional jobs is linear; on the average, HVM gets a 12.28x speedup by passing from 1 to 20 parallel jobs. Furthermore, HVM significantly improves performance without sacrificing optimization accuracy. To show this point, Table 3 reports the comparison between the execution time of HVM and simple binary search. In these experiments, we assumed a budget of 50 units. The results shown in Table 3 conclude that HVM achieves an average speed-up in computation performance of 43% with respect to simple binary search.

## 5 RELATED WORK

Significant research efforts have been dedicated to investigate the incentivization issue in the broader area of *crowdsourcing*, which is defined as the process of obtaining needed services, ideas, or content by soliciting contributions from a large group of people, and especially from an online community, rather than from traditional employees or suppliers.

Zhang and van der Schaar proposed in [23] reputation-based incentive mechanisms for crowdsourcing, where participants will earn reputations upon the completion of a task. Kamar and Horvitz designed incentive mechanisms for consensus tasks that have correct answers to incentivize users for reporting true information [24]. They introduced a novel payment rule, called consensus prediction rule, for evaluating the users' reports in order to determine the payments. Nath *et al.* focused on incentive mechanisms design to minimize the total cost or minimize the total time for executing the task [25]. However, they considered sybil proofness, budget balance, contribution rationality, collapse-proofness, but not truthfulness. In [26], Singla and Krause presented a novel, no-regret posted price mechanism in stochastic online settings. Using a different approach, Goel *et al.* proposed in [27] a truthful mechanism, TM-Uniform, for crowdsourcing markets with a budget constraint. They proved that TM-Uniform is budget feasible, individually rational, truthful, and is 3-approximate compared to the optimum solution. However, they assumed that each user is only allowed to work on one task. Therefore, the task assignment process is basically a matching between tasks and users. Incentive mechanism design was also studied for problems more related to the networking field, such as spectrum trading [28], and cooperative communications [29], among others.

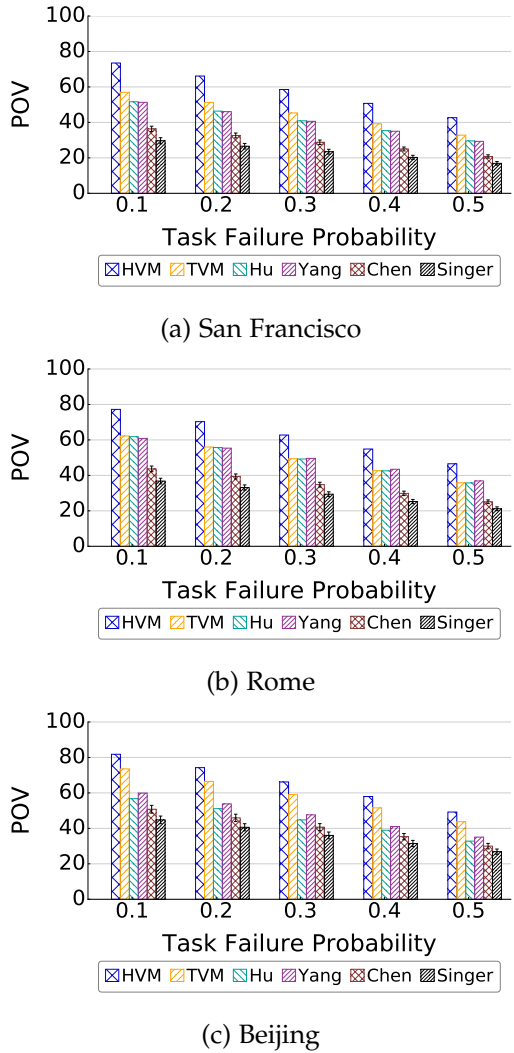

Fig. 10: POV vs. Task Failure Probability.

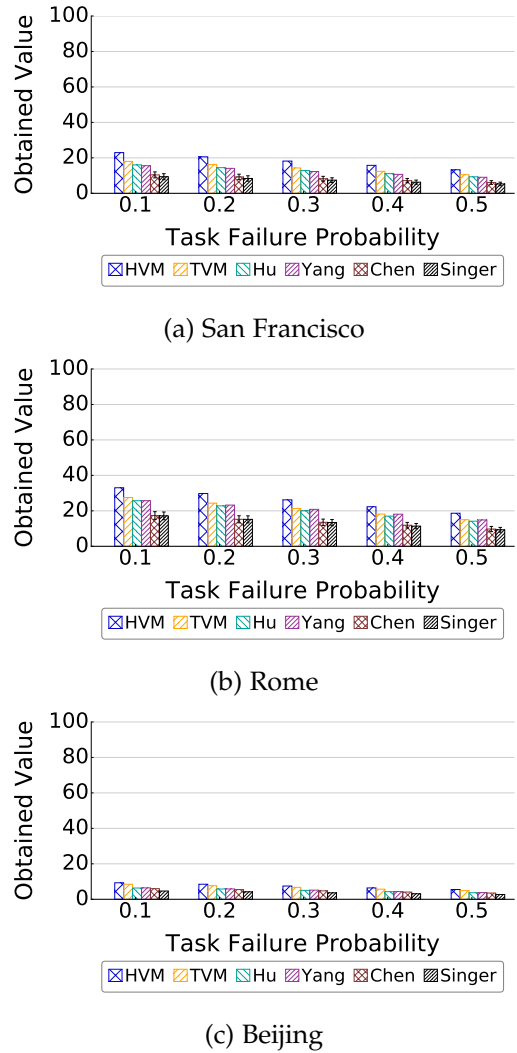

Fig. 11: Obtained Value vs. Task Failure Probability.

The design of incentive mechanisms for mobile crowdsensing systems is particularly challenging, and differentiates from the aforementioned forms of incentive mechanism design. Specifically, as participants move over the sensing area and submit reports through their mobile devices, the value of their contributions for the mobile crowdsensing system is strongly dynamic as it depends from the current location of the participant. For example, a traffic report submitted from a not particularly important location or delayed by several minutes might not be of much value for the system. The selection and reward assignment processes must therefore reflect this aspect, which has been neglected in most of existing mechanisms.

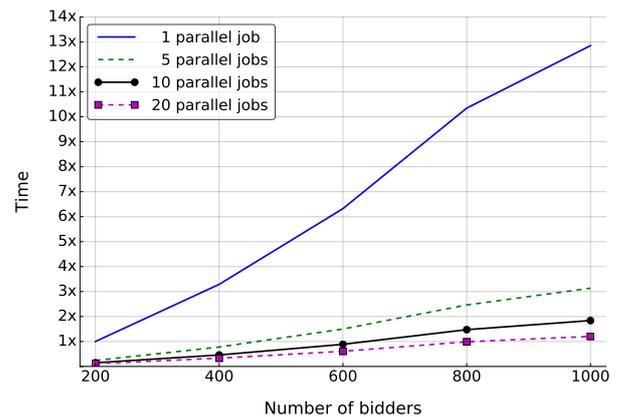

Fig. 12: Execution time vs. Number of bidders vs. Number of parallel jobs.

In recent years, incentive mechanisms in mobile crowdsensing have been widely studied – for excellent surveys on the topic, the reader may refer to [7], [8]. Among these mechanisms, a clear distinction is made based on the *methodology* being used to incentive participation. In particular, we differentiate between *game-theoretical* approaches, which are grounded in mathematical models that provide a foundation for reasoning about rational decision-making, and mechanisms that exploit different concepts, such as micropayments [30]. Game-theoretical approaches can be further categorized, distinguishing between mechanisms that exploit auction theory [31] and mechanisms that instead use other concepts such as Stackelberg games [18].

Most of the existing incentive mechanisms for mobile crowdsensing based on auction theory focus on either maximizing the total utility/value of the platform under a certain constraint (e.g., budget) or minimizing the total payment of the platform. Among the earliest works on the more general field of budget-feasible auction mechanisms are [12] and [13], where the authors proposed randomized mechanisms achieving approximation factors of $\approx 0.008$ and $0.1264$. In the narrower scope of mobile crowdsensing, one of the first mechanisms was presented in [32], where Jaimes *et al.* propose a mechanism that addresses the coverage problem with budget constraints (shown to be NP-Hard) by greedily finding a set of users that covers the greatest possible area within a budget constraint. However, the mechanism fails to consider truthfulness. The problem of guaranteeing a truthful incentive mechanism was explored for the first time in [33]. In this paper the authors propose a model where the system announces a set of sensing tasks, each one having a certain value to the system. Each user then selects a subset of tasks according to its preference and bids for each of them. The system then selects the participants so as to maximize a submodular value function. Although the mechanism shows important properties and achieves good performances, it does not consider budget-feasibility, participants' location nor mobility pattern.

Recently, in [34], the authors consider the sensor selection problem taking the long-term user participation incentive into explicit consideration. To this end, they propose a VCG auction policy for the on-line sensor selection, which achieves a constant competitive ratio of $\mathcal{O}(1)$ with the optimal offline solution. In [35], the authors develop an adaptive and distribute algorithm OptMPSS to maximize phone user financial rewards accounting for their costs. To incentivize phone users to participate they propose an algorithm that merges stochastic Lyapunov optimization with mechanism design theory. The authors show that the mechanism achieves an optimal gross profit for all phone users. Feng *et al.* proposed in [22] an incentive mechanism which takes into account the location of the smartphone users. Specifically, the tasks here are location-based, and participants can bid only on tasks which are in the sensing coverage of the smartphone. The main limitation of [22] is that the sensing tasks and users' positions are assumed to be known in advance and static. A similar mechanism considering dynamic tasks and users has been proposed by the same authors in [36]. However, in both of these works, the budget-feasibility has not been considered. A number of frameworks aimed to recruit participants in order to maximize the coverage of the sensing area have been recently proposed [37]–[42]. In [37], the authors propose a framework to ensure coverage of the collected data, localization of the participating smartphones, and overall energy efficiency of the data collection process. Zhang *et al.* proposed in [38] *CrowdRecruiter*, a framework that minimizes incentive payments by selecting a small number of participants.

## 6 CONCLUSIONS

In this work, we have proposed IncentMe, a framework to incentivize the participation of users with uncertain mobility in mobile crowdsensing. First, we have defined the IncentMe architectural model and formulated the Budgeted Value Maximization (BVM) problem, and proved that it is NP-Hard. By observing that its objective function is submodular, we proposed two mechanisms that achieve proven approximation ratio, as well as satisfy the desired auction-theoretical property of truthfulness and individual-rationality. We have evaluated our mechanisms through experimental study and compared them with the relevant existing work. Results have shown that our algorithms outperform existing methods by almost 30%.

**Francesco Restuccia** (M'16) received his Ph.D. in Computer Science from Missouri S&T in December 2016 under Prof. Sajal K. Das. Currently, he is a Postdoctoral Research Associate with Northeastern University, Boston, MA, USA. His research interests include pervasive and mobile computing and the Internet of Things. He is a Member of the IEEE.

**Pierluca Ferraro** received his Ph.D. in Computer Engineering from the University of Palermo, Italy in 2017. His current research interests include mobile and pervasive computing, Wireless Sensor Networks, and Ambient Intelligence.

**Simone Silvestri** (M'11) graduated with honors and received his PhD in computer science at Sapienza University of Rome, Italy. He is now an Assistant Professor at the Computer Science Department of the University of Kentucky. His research interests lie in the area of network management, hybrid wireless sensor networks, interdependent cyber-physical systems, and smart grid security. He is a Member of the IEEE.

**Sajal K. Das** (F'15) is the Daniel St. Clair Endowed chair in computer science at the Missouri University of Science and Technology. He has published more than 600 research articles in high quality journals and refereed conference proceedings. Prof. Das is the founding Editor-in-Chief of the Pervasive and Mobile Computing journal, and an Associate Editor of the IEEE Transactions on Mobile Computing and ACM Transactions on Sensor Networks. He is a Fellow of the IEEE.

**Giuseppe Lo Re** (SM'11) is an Associate Professor of Computer Engineering at the University of Palermo. His current research interests are in the area of computer networks and distributed systems, broadly focusing on Wireless Sensor Networks, Ambient Intelligence, Internet of Things. He is a Senior Member of the IEEE.